\documentclass[12pt]{iopart}
\usepackage{amssymb}
\usepackage{tikz}
\usepackage[mode=buildnew]{standalone}
\usepackage{caption}
\usepackage{subcaption}
\usepackage{nameref}
\usepackage{censor}
\usepackage{siunitx}

\begin{document}
\title[A framework for comparative analysis with channel attention mechanisms]{EEG motor imagery decoding: A framework for comparative analysis with channel attention mechanisms}

\author{Martin Wimpff\textsuperscript{1}, Leonardo Gizzi\textsuperscript{2}, Jan Zerfowski\textsuperscript{3}, Bin Yang\textsuperscript{1}}

\address{\textsuperscript{1}Institute of Signal Processing and System Theory, University of Stuttgart, Germany\newline\textsuperscript{2}Fraunhofer Institute for Manufacturing Engineering and Automation IPA, Germany\newline\textsuperscript{3}Clinical Neurotechnology Laboratory, Department of Psychiatry and Neurosciences, Charit\'{e} Campus Mitte (CCM), Charit\'{e} - Universitätsmedizin Berlin, Germany}
\ead{martin.wimpff@iss.uni-stuttgart.de}

\begin{abstract}
\textit{Objective}
The objective of this study is to investigate the application of various channel attention mechanisms within the domain of brain-computer interface (BCI) for motor imagery decoding. Channel attention mechanisms can be seen as a powerful evolution of spatial filters traditionally used for motor imagery decoding. This study systematically compares such mechanisms by integrating them into a lightweight architecture framework to evaluate their impact. 
\textit{Approach}
We carefully construct a straightforward and lightweight baseline architecture designed to seamlessly integrate different channel attention mechanisms. This approach is contrary to previous works which only investigate one attention mechanism and usually build a very complex, sometimes nested architecture. Our framework allows us to evaluate and compare the impact of different attention mechanisms under the same circumstances. The easy integration of different channel attention mechanisms as well as the low computational complexity enables us to conduct a wide range of experiments on four datasets to thoroughly assess the effectiveness of the baseline model and the attention mechanisms.  
\textit{Results}
Our experiments demonstrate the strength and generalizability of our architecture framework as well as how channel attention mechanisms can improve the performance while maintaining the small memory footprint and low computational complexity of our baseline architecture.
\textit{Significance}
Our architecture emphasizes simplicity, offering easy integration of channel attention mechanisms, while maintaining a high degree of generalizability across datasets, making it a versatile and efficient solution for EEG motor imagery decoding within brain-computer interfaces.
\end{abstract}
\section{Introduction}
In the field of modern neuroscience and technology, researchers and clinicians are exploring innovative approaches to bridge the gap between the human brain and external devices. One avenue of research is electroencephalogram (EEG) motor imagery decoding, which holds great promise for improving neurorehabilitation strategies \cite{khan2020review}. EEG has emerged as a key technology for deciphering the intricate relationship between the mind and bodily movements because it is a non-invasive, versatile and affordable technique for measuring brain activity. Motor imagery, the mental simulation of movement without actual execution, generates EEG patterns similar to those generated during actual execution. These patterns can be decoded and translated into meaningful commands, enabling direct communication between the brain and external devices. Brain-computer-interfaces (BCIs), based on the principles of EEG motor imagery decoding, provide a pathway for individuals with motor impairments to regain lost functionality. Such interfaces offer unprecedented opportunities for people with conditions like spinal cord injuries, stroke or neurodegenerative diseases to restore a degree of independence and improve their quality of life.
One critical aspect of BCIs is the decoding algorithm used to translate the measured brain activity into meaningful commands. Our work is centered on understanding and investigating different deep learning solutions for the decoding problem.\newline
Since the introduction of AlexNet in 2012, deep neural networks (DNN) have transformed many fields including computer vision, natural language processing and medicine. Throughout the years, the underlying architectures became bigger, wider and deeper to solve increasingly complex tasks at the level of a human and beyond. Currently, generative pre-trained transformer (GPT) models like ChatGPT as well as vision transformers are creating a new hype around the role of attention and especially around the transformer architecture. While the attention mechanism and transformers are well understood and investigated in the natural language processing and computer vision area, attention is still largely unexplored in the context of BCI and EEG. 
\newline
\cite{liu2022sincnet, li2021temporal, altuwaijri2022multi, jia2021mmcnn, zhang2021motor, sun2020eeg} all use a squeeze-and-excitation (SE) module in their architecture. 
\cite{liu2022sincnet} builds a complex hybrid architecture with common spatial patterns (CSP), SincNet filters, an SE module, temporal and spatial convolutions as well as gated recurrent units (GRU). \cite{li2021temporal} created a feature fusion model that uses the SE block to combine deep and multi-spectral features. \cite{altuwaijri2022multi, jia2021mmcnn} created multi-branch convolutional neural networks (CNN) in combination with SE modules. The architecture by \cite{jia2021mmcnn} additionally involves an Inception-Block \cite{szegedy2017inception} and a ResNet \cite{he2016resnet}.  \cite{zhang2021motor} and \cite{sun2020eeg} use time-frequency representations instead of raw time series. 
\newline
\cite{amin2021attention, song2022eeg, altaheri2022physics, ingolfsson2020tcn, zhang2020graph, ma2022time, miao2023lmda, liu2022tcacnet, tao2020eeg} use other types of attention mechanisms.
\cite{amin2021attention} uses an inception-like CNN together with a kind of attention that is not described further. We believe that it is something within the category of multi-head self-attention (MHSA) \cite{vaswani2017attention}. \cite{song2022eeg} built an architecture consisting of a convolutional stage as in ShallowNet \cite{schirrmeister2017} and EEGNet \cite{lawhern2018eegnet} followed by MHSA blocks attending over the temporal direction. \cite{altaheri2022physics} uses a similar convolutional stage followed by an ensemble of blocks consisting of MHSA and temporal convolutional blocks \cite{ingolfsson2020tcn}. \cite{zhang2020graph} on the other hand uses a graph-based CNN together with a recurrent attention module composed of long short-term memory cells (LSTM). \cite{ma2022time} uses a complex time-distributed attention architecture, where the EEG signal is initially split into non-overlapping segments and further spatially filtered. Afterwards, the segments get classified through parallel attention modules (one for each segment) followed by an LSTM and a fully connected layer.
\newline
Further, there are some works that combine different attention mechanisms like \cite{miao2023lmda, liu2022tcacnet, tao2020eeg}. \cite{miao2023lmda} combines a custom lightweight channel recalibration module with the channel attention module from ECA-Net \cite{wang2020eca}. \cite{liu2022tcacnet} combines a channel attention module based on the wavelet packet sub-band energy ratio with a temporal attention mechanism followed by a feature fusion architecture. \cite{tao2020eeg} builds a model for emotion recognition that also uses a form of channel-wise attention together with LSTMs and self-attention.
We observed that while there is a number of publications about EEG-based BCI using attention,
most of them use very complex and task-specific, computationally inefficient, hybrid/multi-branch/multi-scale/ensemble/time-distributed architectures and all of them investigate only one attention mechanism. 
The two time-frequency solutions \cite{zhang2021motor, sun2020eeg} as well as the EEGConformer \cite{song2022eeg} are an exception by having a simple, non-nested architecture.
\newline
The insights learned from such specialized and narrow investigations as well as the ability to transfer these learnings to other problems are often limited.
An additional problem with complex architectures is their computational complexity as well as comparability. Different models are difficult to compare as each of them uses different layers before and after the layers with attention as well as a different training routine. Those are all reasons that lead to high entry barriers, confusion and mistrust by medical experts towards the deployment of deep learning in BCI. Hence, most of the experts still rely on classical algorithms like CSP \cite{blankertz2007csp, ang2008fbcsp} or on the somewhat outdated but well explored shallow architectures EEGNet or ShallowNet.
\newline
To overcome these limitations, we first introduce a simple yet powerful baseline model without any attention mechanism. We will also showcase the process of how we developed this baseline architecture from the well-known EEGNet and ShallowNet to justify our design choices. 
We then integrate different channel attention mechanisms into this architecture framework while keeping the network and training routine constant across all experiments to allow for a fair comparison. This allows us to measure the real impact and suitability of each attention mechanism.
We believe that this fair and systematic comparison is able to increase the understanding of attention in BCIs. Moreover, it serves as a foundational platform for forthcoming research in the realms of attention in BCI. The source code is available at https://github.com/martinwimpff/channel-attention and was developed using the python packages pytorch \cite{Paszke_PyTorch_An_Imperative_2019}, pytorch-lightning \cite{Falcon_PyTorch_Lightning_2019}, braindecode \cite{schirrmeister2017}, mne-python \cite{mne-python} and moabb \cite{jayaram2018moabb}.

\section{Method}
\subsection{Task and datasets}
In this study, we will utilize four distinct datasets across two different settings. This aims to encompass a broad spectrum of motor imagery dataset characteristics while also exploring the potential boundaries of our methodology.
We will use the 2a \cite{brunner2008bci} and 2b \cite{leeb2008bci} dataset from the BCI competition IV (BCIC IV) and the High Gamma dataset (HGD) \cite{schirrmeister2017}. Additionally we will use the imbalanced dataset IVa from the BCI competition III \cite{blankertz2006bci}.
The first three datasets are frequently employed in deep learning research and exhibit a range of sensor quantities, spanning from 3 to 44 sensors. Conversely, the BCIC III IVa dataset is typically decoded using traditional methods, given that the training trials can be as minimal as 28 in number.
This variety of datasets enables a thorough investigation, allowing for effective comparison and contrast of different methodologies. In the following we will briefly summarize the different datasets.\newline
Both datasets from the BCIC IV consist of data from 9 different subjects and were recorded at \SI{250}{\hertz}. The 2a dataset was captured using 22 electrodes and consists of trials imagining one of four distinct movements (feet, left hand, right hand and tongue movement).
Per subject, two sessions a 288 trials each lasting 4 seconds were acquired. The trials of the first session will be used for training, the trials from the second session will be used for testing.\newline
The 2b dataset only uses three electrodes to record the data from two different imagined movements (left hand, right hand). Additionally, the trials last 4.5s instead of 4s and the data was recorded in 5 sessions of which the first three (400 trials) will be used for training and the last two (320 trials) will be used for testing. Thus the first three sessions of the 2b dataset correspond to the first session of the 2a dataset and the last two sessions of the 2b dataset correspond to the second session of the 2a dataset, respectively.\newline
The HGD is the biggest dataset used in this investigation and has EEG data from 14 subjects recorded at \SI{250}{\hertz} on 128 electrodes of which only 44 sensors covering the motor cortex are used. Similar to the 2a dataset, the trials lasted 4s and four different movements (left hand, right hand, feet, rest) were imagined. The first session consists of approx. 880 trials per subject and will be used for training whereas the second session consists of approx. 160 trials per subject and will be used for testing.\newline
To investigate our method for small datasets we also employ experiments with the BCIC III IVa dataset which involves data from five subjects. This dataset has 280 trials per subject recorded at \SI{100}{\hertz} using 118 electrodes and consists of trials from two distinct classes (right hand, foot). What differentiates this dataset from the others is the imbalanced split into trials used for training and trials used for testing. The number of training trials ranges from 28 trials to 224 trials depending on the subject (aa: 168, al: 224, av: 84, aw: 56, ay: 28) and the remaining trials are used for testing. Each trial lasted 3.5s. To reduce the number of data points per trial to prevent overfitting we selected the three channels shared between the BCIC IV 2b dataset and the BCIC III IVa dataset (C3, Cz and C4). \newline
We investigate two different scenarios, within-subject classification and cross-subject leave-one-subject-out classification. For both settings we train one model per subject and use the data from the second session of the target subject for testing.
For the within-subject classification, the data from the first session of the target subject is used for training such that the model is trained and tested on data from the same subject.
In the cross-subject scenario, the model is trained using data from the first session of all subjects except the target subject, aiming to assess its generalization capability to previously unseen subjects. \newline
As we employ single-trial classification we always use the full trial at the original sampling frequency to allow a fair comparison to other approaches. 
However, for the BCIC IV 2b dataset, we only utilize the initial 4 seconds of the entire 4.5-second trial, a practice more commonly observed in the literature. 
\subsection{BaseNet}
Previous publications \cite{schirrmeister2017, lawhern2018eegnet, li2021temporal, altuwaijri2022multi,song2022eeg, altaheri2022physics, ingolfsson2020tcn} working on EEG motor imagery decoding have shown that it is helpful to use a stem-block as a first layer of the architecture. This block is usually composed of a temporal convolutional layer and a spatial convolutional layer along with normalization layers, nonlinearity and pooling layers. Since EEGNet and ShallowNet, most of the successive deep learning architectures used a somewhat similar approach. The problem, however, is that every publication tends to use a slightly different stem-block, some are closer to the one used in ShallowNet, some are more similar to the one in EEGNet. 
Different kernel sizes, number of filters, normalizations, nonlinearities, pooling layers and weight initializations between different architectures impair the comparability as it is not clear to which part of the network a possible improvement can be attributed to. Moreover, comparing a very complex and specialized architecture to a somewhat dated architecture like EEGNet or ShallowNet without adjusting the hyperparameters and training routine raises concerns about the validity of the comparison.
To overcome this limitation, we decided to develop a unified shallow neural network called BaseNet, which aims to be parameter-efficient and expressive with plausible and simple hyperparameter choices.
To do so, we gradually combined, simplified and modernized the design choices from EEGNet and ShallowNet to create a strong baseline model. The architecture is visualized in Figure \ref{fig:basenet}.\newline
The first step is the simplification of the training routine. We use the full 4s trial at \SI{250}{\hertz} instead of a shortened trial at a lower sampling frequency to provide the full original information to the network to allow a fair comparison to other approaches. Further, we train the model for a fixed number of epochs using a learning rate scheduler with a linear warmup followed by a cosine decay instead of using early stopping to reduce loss oscillations while avoiding an additional data split.\newline
The second step involves the simplification of the architecture. To simplify, we use a linear classification layer and always use the default initialization and batch normalization of pytorch. Further, our model does not use bias in the convolutional layers and does not use any kernel weight constraints. We use the grouped convolutions from EEGNet as well as their intermediate batch normalization layer between the first two layers to reduce the number of parameters and to regularize the training.\newline
The next step involves the choice of activation function, kernel sizes and pool sizes. We chose the ELU activation function from EEGNet instead of the square-log function from ShallowNet as it resulted in better performance. We used the kernel and pool sizes of ShallowNet as it was originally developed for trials at \SI{250}{\hertz} in contrast to EEGNet, which was developed for trials at \SI{128}{\hertz}. 
We chose to use the depthwise separable convolution from EEGNet to add depth to the model while keeping a low memory footprint. To decouple the channel dimension in the spatial layer from the channel dimension in the depthwise separable convolution, we further added a $1\times1$ convolutional channel projection layer as in EEGConformer \cite{song2022eeg}. We then used the filter sizes from EEGNet for the depthwise separable convolution and the filter sizes from ShallowNet for the first two convolutional layers.
\newline
To investigate the use of the different channel attention mechanisms, the attention mechanisms introduced in the next section have to be integrated into the BaseNet architecture. We do this by simply adding the channel attention block between the nonlinearity of the depthwise separable convolution and the pooling layer. This position is visualized by a red dot in Figure \ref{fig:basenet}.

\subsection{Attention}
Attention enables humans to selectively process and prioritize information from a multitude of sensory stimuli, focusing on those that are relevant to specific goals and cognitive processes. The deep learning community has tried to include different kinds of attention mechanisms for many years (see \cite{guo2022attentionreview} for a review). 
The different attention mechanisms can be macroscopically categorized by the domain they are operating in. \cite{guo2022attentionreview} distinguishes between channel, spatial, temporal and branch attention as well as combinations where the attention mechanism operates on two domains. \\
The success and wide use of spatial filters for motor imagery decoding highlights the importance of the spatial distribution of brain activity. The areas of interest for motor imagery decoding (namely, the motor and sensory cortices) are somatotopically organized, and different cortical areas can be associated with different functions or areas of the body. We will therefore focus on the spatial dimension and compare different channel attention mechanisms in this paper. 
As EEG data and images have different dimensions, we first want to clarify the terms "spatial" and "channel" to avoid further confusion. 
\newline
EEG data typically consists of multiple time series (temporal dimension) from different sensors (spatial dimension) and can be described by a tensor \(X_{eeg} \in \mathbb{R}^{C \times T}\) where $C$ is the number of EEG sensors and $T$ is the number of time points. Images, on the other hand, are three-dimensional tensors \(X_{img} \in \mathbb{R}^{C \times H \times W}\) where $C$ describes the number of color channels of the input image and later the number of feature channels of intermediate tensors, $H$ the height and $W$ the width. As we investigate the use of channel attention mechanisms originating from the image domain for EEG data, the dimension of the color channels $C$ from images \(X_{img}\) corresponds to the spatial dimension ($C$) of the EEG tensor \(X_{eeg}\). We also examine two architectures with additional attention in the spatial domain in images ($H$ and $W$) which transfers to the temporal domain $T$ in EEG data. 
It is worth noting that the attention mechanisms attend over feature channels rather than EEG channels as the second layer (spatial convolution) as well as the channel projector already filter the original input spatially. This is similar to images where the color channels become feature channels after the first convolutional layer.
In the following sections we will only use the terms "channel" and "temporal" for clarity.
\subsubsection{General form}
The general process of attending to a specific region can be formulated as 
\begin{equation}
    Attention = f(a(x), x)
\end{equation}
where $a(x)$ represents the function that generates the attention. $f(a(x), x)$ means that the input $x$ is processed based on the attention generated by $a(x)$ (cf. \cite{guo2022attentionreview}). This general form allows us to describe different attention mechanisms in a unified manner.
\subsubsection{Squeeze-and-excitation}
\begin{figure}
    \centering
    \begin{subfigure}[b]{0.22\textwidth}
        \centering
        \includegraphics{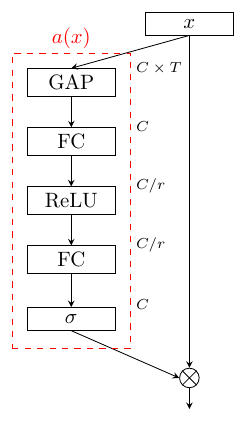}
        \caption{SE Block}
        \label{ch-att:se}
    \end{subfigure}
    \hfill
    \begin{subfigure}[b]{0.22\textwidth}
        \centering
        \includegraphics{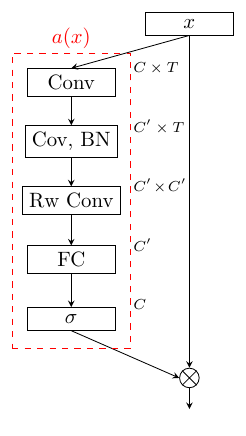}
        \caption{GSoP Block}
    \end{subfigure}
    \hfill
    \begin{subfigure}[b]{0.22\textwidth}
        \centering
        \includegraphics{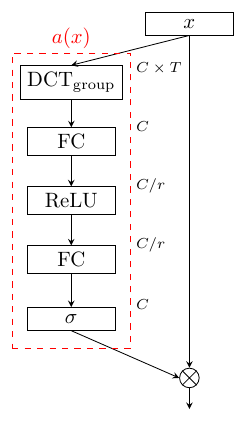}
        \caption{FCA Block}
    \end{subfigure}
    \hfill
    \begin{subfigure}[b]{0.22\textwidth}
        \centering
        \includegraphics{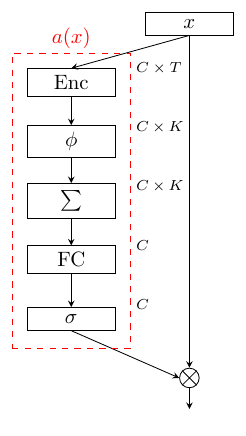}
        \caption{EncNet Block}
    \end{subfigure}
    \hfill
    \begin{subfigure}[b]{0.22\textwidth}
        \centering
        \includegraphics{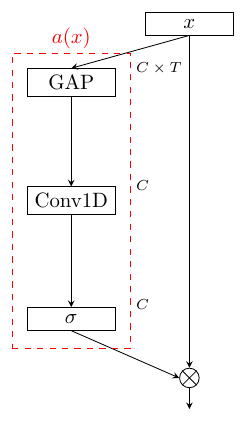}
        \caption{ECA Block}
    \end{subfigure}
    \hfill
    \begin{subfigure}[b]{0.22\textwidth}
        \centering
        \includegraphics{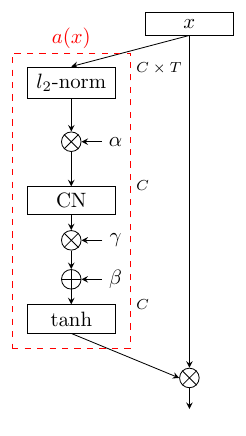}
        \caption{GCT Block}
    \end{subfigure}
    \hfill
    \begin{subfigure}[b]{0.22\textwidth}
        \centering
        \includegraphics{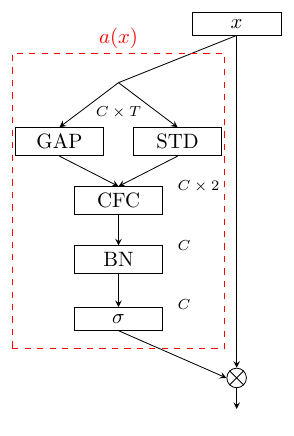}
        \caption{SRM Block}
    \end{subfigure}
    \caption{Channel attention mechanisms. GAP = global average pooling, FC = fully-connected layer, Cov=covariance pooling, BN=batch normalization, RW Conv = row-wise convolution, $\text{DCT}_\text{group}$=grouped discrete cosine transform, Enc=encoder block, CN=channel normalization, STD = global standard deviation, CFC= channel-wise fully-connected layer. Adapted from \cite{guo2022attentionreview}.}
    \label{fig:channel-attention}
\end{figure}

Squeeze-and-excitation network (SENet) \cite{hu2018squeeze} is a simple yet powerful way to enhance feature representations by adaptively re-calibrating channel-wise information. SENet can be seen as the pioneer of channel attention on which most of the subsequent works are founded.
It can be formulated as 
\begin{equation}
    a(x) = \sigma(\text{MLP}(\text{GAP}(x))) = \sigma(W_2\phi(W_1\text{GAP}(x))) 
\end{equation}
\begin{equation}
    f(a(x), x) = a(x)\otimes x
\end{equation}
where global average pooling (GAP) calculates the average of $T$ samples for each channel (\(\mathbb{R}^{C\times T}\rightarrow\mathbb{R}^{C}\)) and the multi-layer-perceptron (MLP) learns adaptive attention weights (\(\in \mathbb{R}^{C}\)) per channel. The MLP consists of two fully connected Layers (FC) connected by a ReLU activation function ($\phi$) to enable nonlinearity. The first FC projects the channel dimension $C$ to a lower dimension $C'=C/r$ by multiplying with $W_1\in\mathbb{R}^{C'\times C}$, where $r\geq1$ is called the reduction rate. The second FC projects the attention weights back to the initial channel dimension $C$ by multiplying with $W_2\in\mathbb{R}^{C\times C'}$ followed by a sigmoid activation function $\sigma$ mapping each value to a range between 0 and 1. Finally, each channel of the original input $x$ gets multiplied elementwise by $a(x)$. The GAP layer represents the squeeze module, whereas the excitation module comprises of the MLP and the sigmoid function. 
The SE block is visualized in Figure \ref{ch-att:se}.
The following channel attention mechanisms try to improve the SE block in different ways. We arranged the following sections based on the improved part of SE. All channel attention mechanisms studied in this paper are visualized in Figure \ref{fig:channel-attention}.
\subsubsection{Improving the squeeze module}
\cite{gao2019gsop, qin2021fcanet, zhang2018context} try to improve the original SE module by changing the squeeze module which is a GAP layer in SE to something more expressive. \cite{gao2019gsop} argue that a GAP layer is limited because it only calculates first-order statistics. Therefore they introduce a global second-order pooling (GSoP) block to model second-order statistics. 
Mathematically, the attention part of a GSoP block is given by
\begin{equation}
    a(x) = \sigma(\text{FC}(\text{RwConv}(\text{Cov}(\text{Conv}(x)))))
\end{equation}
The initial $1\times1$ convolution reduces the number of channels $C$ to $C'$. Afterwards, a $C'\times C'$ covariance matrix is computed. Subsequently, a row-wise normalization is performed through a row-wise convolution which results in a one dimensional feature vector $\in\mathbb{R}^{C'}$. The fully-connected layer as well as the sigmoid function is similar to SE.
\cite{qin2021fcanet}, on the other hand, view the squeeze module as a compression problem and use the discrete cosine transform (DCT) to decompose the features in the frequency domain. The GAP layer of SE is replaced by a grouped DCT module that divides the original input with $C$ channels into $n_{groups}$ groups with each $C'=C/n_{groups}$ channels and then multiplies each group of channels element-wise with different DCT bases corresponding to different frequencies. $\text{DCT}^0$ is the DCT base with lowest frequency component and is equivalent to a scaled GAP layer (cf. \cite{qin2021fcanet} for mathematical derivation). The MLP and the sigmoid function stay the same as in SE. Formally, the attention of their Frequency Channel Attention Network (FCA) can be formulated as
\begin{equation}
    a(x) = \sigma(\text{MLP}(\text{DCT}(\text{Group}(x))))
\end{equation}
\cite{zhang2018context} proposed a method particularly suited to semantic segmentation. Their attention mechanism has $K$ learnable codewords or visual centers $d_i\in\mathbb{R}^C$ with corresponding learnable smoothing factors $s_i\in\mathbb{R}$. Soft-assignment weights $e_k=\sum_{i=1}^Te_{ik}$ are calculated with
\begin{equation}
    e_{ik} = \frac{\text{exp}(-s_k||r_{ik}||^{2})}
    {\sum_{j=1}^K\text{exp}(-s_j||r_{ij}||^{2})}r_{ik}, \qquad r_{ik} = x_i - d_k, \qquad x_i\in\mathbb{R}^C
\end{equation}
The soft-assignment weights are further aggregated by $e=\sum_{k=1}^K\phi(e_k)$ where $\phi$ represents a batch normalization and a ReLU activation function. The final attention function is then given by $a(e)=\sigma(\text{FC}(e))$. They additionally use a second loss function especially tailored for semantic segmentation which we will not use, as we don't do semantic segmentation and only have one class per trial.
\subsubsection{Improving the excitation module}
The ECA-Net \cite{wang2020eca} is the only method in this comparison that soley relies on improving the excitation module. In the original SE module the excitation module consists of two FC layers with $C^2/r$ parameters each. To reduce the number of parameters, they proposed a method called local cross-channel attention that tries to balance the trade-off between cross-channel interaction and parameter efficiency by using a 1D convolution in the channel dimension after the GAP layer. The attention mechanism of ECA-Net can be formulated as 
\begin{equation}
    a(x) = \sigma(\text{Conv1D}(\text{GAP}(x)))
\end{equation}
which only needs $k$ parameters where $k$ is the kernel size of the 1D convolution.
\subsubsection{Improving the squeeze and the excitation module}
An early follow-up by the original authors called gather-excite (GE) \cite{hu2018gather} investigated the parameterization of the squeeze and the deparameterization of the excitation module. Originally, the squeeze module (GAP layer) has zero parameters. GE proposes a gather module that gathers information through a grouped convolution with a variable extent. If the kernel size $k$ equals the length of the sequence ($k=T$), the extent is called global. A global gather module introduces $Ck=CT$ new parameters. They further investigate a parameter-free excitation module that simply scales each channel based on the sigmoid of its average value without any MLP ($a(x)=\sigma(\text{GAP}(x))$).
\newline
Another method trying to reduce the number of parameters is the gated channel transformation (GCT) \cite{yang2020gated}. This method aggregates channel information by computing the $l_2$-norm per channel instead of the average. There are three learnable parameters $\alpha_c, \beta_c$ and $\gamma_c$ per channel $c$ controlling the gating mechanism. Instead of a sigmoid function, $\tanh$ is used. GCT is given by
\begin{equation}
    s_c = \alpha_c\cdot ||x_c||_2, \quad \hat{s}_c = \frac{\sqrt{C}s_c}{[(\sum_{c=1}^Cs_c^2)+\epsilon]^{\frac{1}{2}}}, \quad \epsilon=10^{-5}
\end{equation}
\begin{equation}
    a(x) = 1 + \tanh(\gamma\hat{s}+\beta)
\end{equation}
which reduces the number of parameters from $2C^2/r$ (SENet) to $3C$. 
\newline
While the previous approaches (except the parameter-free version from GE) all rely on cross-channel interaction, the style-based recalibration module (SRM) \cite{lee2019srm} investigates a channel-independent attention mechanism. They first collect $d=2$ style features (average and standard deviation) per channel, which they call style pooling. The following step is called style integration which linearly combines the style features per channel through a channel-wise fully connected (CFC) layer.
The attention mechanism can be formulated by 
\begin{equation}
    a(x) = \sigma(\text{BN}(\text{CFC}(\text{Concat}([\text{GAP}(x), \text{STD}(x)]))))
\end{equation}
where BN is a batch normalization layer and the CFC layer has $dC$ learnable parameters.
\subsubsection{Channel and temporal attention mechanisms}
SENet and the subsequent approaches used the attention mechanism only in the channel domain. However, there are combinations which attend over the temporal and channel domain.
\newline
The pioneering work for combining both types of attention is the convolutional block attention module (CBAM) \cite{woo2018cbam}. CBAM is divided into a channel attention module and a temporal attention module. The channel attention module is almost similar to SENet and the attention mechanism is given by
\begin{equation}
    a_{ch}(x) = \sigma(\text{MLP}(\text{GAP}(x))+\text{MLP}(\text{GMP}(x))) \in \mathbb{R}^C
\end{equation}
where the MLP is shared and the only difference to SENet is the additional (parameter-free) global max pooling (GMP) layer.
Afterwards, nearly the same process is repeated in the temporal domain:
\begin{equation}
    a_{temp}(x) = \sigma(\text{Conv2D}(\text{Concat}([\text{GAP}(x), \text{GMP}(x)]))) \in \mathbb{R}^T
\end{equation}
The GAP and GMP layer now calculate the average and the maximum respectively over all channels, resulting in a concatenated feature map $f\in\mathbb{R}^{2\times T}$. This feature map is then processed by a 2D convolutional layer resulting in a temporal attention map $a_{temp}(x)\in\mathbb{R}^{T}$.
The complete attention mechanism of CBAM is given by 
\begin{equation}
    f(x, a_{ch}(x), a_{temp}(x)) = a_{temp}(a_{ch}(x)\otimes x) \otimes (a_{ch}(x)\otimes x)
\end{equation}
While the temporal and channel attention module in CBAM do not cooperate explicitly, \cite{wu2023cat} introduced a mechanism called CAT specifically designed to allow collaboration between the two types of attention.
In addition to the GAP and GMP layer in CBAM, they use a global entropy layer (GEP). Instead of concatenating the different feature maps, they use a weighted sum to get one feature map per direction of attention
\begin{equation}
    C'_A = C_\alpha\text{MLP}(\text{GAP}(x)) + C_\beta\text{MLP}(\text{GMP}(x)) + C_\gamma\text{MLP}(\text{GEP}(x))
\end{equation}
\begin{equation}
    T'_A = \text{Conv2D}(T_\alpha\text{GAP}(x) + T_\beta\text{GMP}(x) + T_\gamma\text{GEP}(x)
\end{equation}
where the parameters $C_{\alpha/\beta/\gamma}, T_{\alpha/\beta/\gamma}\in \mathbb{R}$ control the collaboration between the aggregated features in the channel attention and temporal attention respectively. The final attention mechanism is then given by
\begin{equation}
    a(x) = \sigma(C_W \cdot C'_A) + \sigma(T_W \cdot T'_A), \quad f(x, a(x)) = a(x) \otimes x
\end{equation}
where $C_W, T_W \in \mathbb{R}$ control the collaboration between the channel and temporal attention.
To investigate the use of collaboration weights for channel attention without temporal attention, we propose a modified version of CAT which we call CATLite. The attention mechanism of CATLite is given by $a(x) = C'_A$.
\subsection{Training}
We train all of our models with the same training routine for each subject.
To preprocess the EEG data, we use a \SI{40}{\hertz} lowpass filter for the BCIC datasets and a \SI{4}{\hertz} highpass filter for the High Gamma dataset. For the other models used in this comparison \cite{li2021temporal, song2022eeg, altaheri2022physics, ingolfsson2020tcn, schirrmeister2017, lawhern2018eegnet} we used the filters they reported for the BCIC datasets while we kept the \SI{4}{\hertz} highpass filter for all experiments with the HGD. We further normalize each channel to have zero mean and unit deviation. 
We then train all models for a fixed number of 1000 epochs and optimize via the Adam optimizer with a learning rate of $10^{-3}$ for the within-subject setting. We further use a learning rate scheduler with a linear warmup of 20 epochs followed by a cosine decay to reduce oscillations of the loss. \newline
For the cross-subject experiments, we train for 125 epochs for the BCIC datasets and for 77 epochs for the HGD as there is more training data. For the BCIC IV datasets, for example, there is eight times more training data available compared to the within-subject setting since each model is trained using data from eight subjects.
For the BCIC III IVa dataset, the amount of training data increases by a factor of 1.5-19 compared to the within-subject setting depending on the subject. 
Therefore, we've decided on a compromise, training all models for 125 epochs, also motivated by the pursuit of consistency. The number of warmup epochs was reduced to 3 and 2 respectively.\newline
The test accuracy is evaluated on the last checkpoint. Each experiment is conducted five times with five different random seeds. The reported average accuracy is the average of these five runs. The reported standard deviation is the standard deviation of the average test accuracy (over all subjects) between these five runs.
This is done because the training process still involves a fair amount of randomness as the datasets are small and noisy and the dropout rate is very high (50\%). 

\section{Results}
\begin{figure}
    \centering
        \begin{subfigure}[b]{0.3\textwidth}
        \centering
        \includegraphics{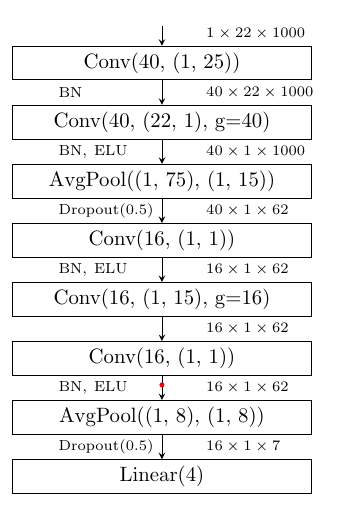}
        \caption{Architecture}
        \label{fig:basenet}
    \end{subfigure}
    \hfill
    \begin{subfigure}[b]{0.65\textwidth}
        \centering
        \includegraphics{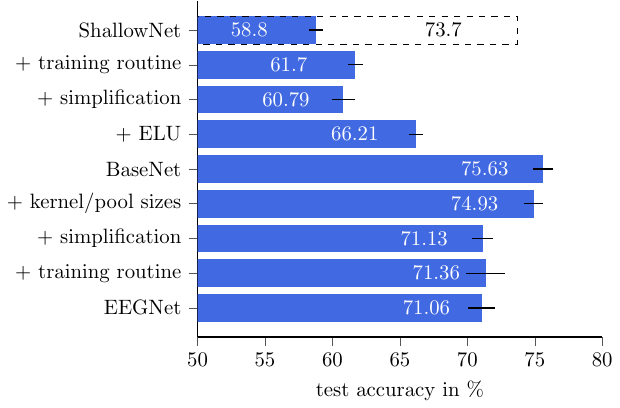}
        \caption{Development}
        \label{fig:basent-development}
    \end{subfigure}
    \caption{Architecture and development of BaseNet. (a): g = number of groups, BN = batch normalization. The red dot indicates the position of the optional channel attention mechanism. (b): Accuracies for the BCIC IV 2a dataset in the within-subject scenario. The dashed bar represents the reported accuracy of ShallowNet with their cropped training strategy. The black error bars indicate the standard deviation for five runs with different random seeds.}
    \label{fig:basent-ablation}
\end{figure}
\begin{table}[]
    \centering
    \caption{Results (in \%) of all models for the BCIC IV 2a dataset in both settings along with their configuration and number of parameters.}
    \label{tab:results-bcic2a}
    \begin{tabular}{ |p{3.7cm}||p{3cm}|r|p{2.6cm}|p{2.5cm}|  }
     \hline
     model & configuration & parameters & within-subject& cross-subject\\
     \hline \hline
     EEGNet \cite{lawhern2018eegnet}  &   -  & 1,716 &   $71.06\pm1.0$ & $52.74\pm 1.34$\\
     ShallowNet \cite{schirrmeister2017} &  -  & 44,644   &$58.8\pm0.51$ & $47.79\pm 0.84$\\
     ShallowNet*\cite{schirrmeister2017} &  cropped training  & 44,644   &73.7 & - \\
     BaseNet & -  & 3,692&  $75.63 \pm 0.75$ & $57.49\pm 0.52$\\
     BaseNet + SE    &r=4 & 3,820&  $76.9\pm0.61$ & $57.55 \pm 0.66$\\
     \hline
     BaseNet + GSoP    &r=4 & 4,120&  $76.28\pm1.05$ & $57.63\pm0.36$\\
     BaseNet + FCA    & r=4, $\text{DCT}^0$ & 4,812&  $76.33\pm0.82$ & $57.44\pm0.6$\\
     BaseNet + EncNet    &4 codewords & 4,040&  $76.37\pm0.76$ & $57.94 \pm0.69$\\
     \hline
     BaseNet + ECANet   & k=9 & 3,695&  $76.86\pm0.63$ & $57.11\pm0.72$\\
     \hline
     BaseNet + $\text{GE-}\theta^-$    & - & 3,692&  $75.57\pm0.35$ & $56.97\pm1.05$\\
     BaseNet + $\text{GE-}\theta$    & e=global & 4,716&  $75.69\pm0.84$ & $56.96\pm0.89$\\
     BaseNet + $\text{GE-}\theta^+$    & e=global, r=4 & 4,844&  $74.95\pm0.87$ & $57.56 \pm0.49$\\
     BaseNet + GCT    & - & 3,740&  $75.96\pm1.04$ & $57.76\pm0.63$\\
     BaseNet + $\text{GCT}_\text{GAP}$    & - & 3,740&  $75.92\pm0.82$& $57.71\pm0.63$\\
     BaseNet + $\text{SE-}l_2$    &r=4 & 3,820&  $75.89\pm0.65$ & $57.28\pm0.73$\\
     BaseNet + SRM    & - & 3,756&  $75.65\pm0.41$&$57.04\pm0.97$\\
     BaseNet + $\text{SRM}_{\text{cross}}$    & r=4 & 4,108&  $76.19\pm0.64$&$56.75\pm0.79$\\
     \hline
     BaseNet + CBAM  & k=15, r=8 & 3,787&  $76.75\pm0.65$ & $57.18\pm0.96$\\
     BaseNet + CAT    & k=3, r=4 & 5,641&  $76.88\pm1.05$ & $\mathbf{57.99\pm1.1}$\\
     BaseNet + CATLite    & r=4 & 3,848 &  $\mathbf{76.91\pm0.9}$ & $57.14\pm0.6$\\
     \hline
    TS-SEFFNet \cite{li2021temporal} & - & 334,824 & $73.5\pm0.42$ & $55.21\pm 0.36$\\
    EEGConformer \cite{song2022eeg} & - & 789,572 & $75.79\pm0.26$ & $44.78 \pm 0.73$\\
    ATCNet \cite{altaheri2022physics} & - & 113,732 & $\mathbf{79.08 \pm 0.43}$ & $\mathbf{57.81 \pm2.0}$ \\
    EEGTCNet \cite{ingolfsson2020tcn}& act=ELU & 4,096&$75.62 \pm 0.66$&$54.51\pm 2.46$\\
     \hline
    \end{tabular}
\end{table}

\begin{table}[]
    \centering
    \caption{Results (in \%) of all models for the BCIC IV 2b dataset in both settings along with their configuration.}
    \label{tab:results-bcic2b}
    \begin{tabular}{ |p{4.5cm}||p{3cm}|p{3cm}|p{3cm}|  }
     \hline
     model & configuration & within-subject & cross-subject \\
     \hline \hline
     EEGNet \cite{lawhern2018eegnet}& - & $83.54\pm0.46$ &   $77.57\pm0.62$\\
     ShallowNet \cite{schirrmeister2017}& - & $78.41\pm0.99$   &$74.63\pm0.56$\\
     BaseNet & - & $84.34 \pm 0.36$&  $78.52 \pm 0.76$\\
     BaseNet + SE  & r=1  & $84.56\pm0.43$&  $\mathbf{78.98\pm0.94}$\\
     \hline
     BaseNet + GSoP    &r=1 &   $81.64\pm0.44$ & $77.13\pm0.75$\\
     BaseNet + FCA    & r=1, $\text{DCT}^0$ & $83.78\pm0.53$ & $77.77\pm1.25$\\
     BaseNet + EncNet    &2 codewords &   $84.33 \pm 0.41$ & $77.87\pm0.66$\\
     \hline
     BaseNet + ECANet  & k=11  & $84.51\pm0.31$&  $78.74\pm0.43$\\
     \hline
     BaseNet + $\text{GE-}\theta^-$    & - &  $84.38\pm0.15$ & $77.94\pm0.42$\\
     BaseNet + $\text{GE-}\theta$    & e=global &   $\mathbf{85.32\pm0.37}$ & $77.68\pm0.71$\\
     BaseNet + $\text{GE-}\theta^+$    & e=global, r=1 &   $83.5\pm0.74$ & $77.64\pm0.76$\\
     BaseNet + GCT    & - &   $84.52\pm0.45$ & $78.47\pm0.79$\\
     BaseNet + $\text{GCT}_\text{GAP}$    & - &  $84.58\pm0.23$& $77.94\pm0.89$\\
     BaseNet + $\text{SE-}l_2$    &r=1 &   $84.21\pm0.39$ & $78.42\pm0.95$\\
     BaseNet + SRM  & -  & $84.84\pm0.69$&  $77.88\pm0.6$\\
     BaseNet + $\text{SRM}_{\text{cross}}$  & r=1  &  $83.13\pm0.52$&  $77.39\pm0.82$\\
     \hline
     BaseNet + CBAM & r=1, k=15 & $84.61\pm0.55$ & $78.58\pm0.49$ \\
     BaseNet + CAT  & r=2, k=9 & $84.96\pm0.44$&  $78.8\pm0.57$\\
     BaseNet + CATLite & r=1   & $84.84\pm0.42$ &  $78.26\pm0.86$\\
     \hline
    TS-SEFFNet \cite{li2021temporal}& - & $83.27\pm0.27$ & $78.15\pm0.3$\\
    EEGConformer \cite{song2022eeg}& - & $81.02\pm0.53$ & $73.41\pm0.52$\\
    ATCNet \cite{altaheri2022physics}& - & $\mathbf{85.52\pm0.44}$ & $78.78\pm0.77$ \\
    EEGTCNet \cite{ingolfsson2020tcn}& act=ELU & $85.51\pm0.42$&$\mathbf{78.83\pm0.46}$\\
     \hline
    \end{tabular}
\end{table}

\begin{table}[]
    \centering
    \caption{Results (in \%) of all models for the HGD in both settings along with their configuration.}
    \label{tab:results-hgd}
    \begin{tabular}{ |p{4.5cm}||p{3cm}|p{3cm}|p{3cm}|  }
     \hline
     model & configuration & within-subject & cross-subject\\
     \hline \hline
     EEGNet \cite{lawhern2018eegnet} & - &   $84.68\pm0.91$& $57.51\pm2.39$\\
     ShallowNet \cite{schirrmeister2017}& - &  $88.72\pm0.85$&$\mathbf{72.47\pm0.94}$\\
     ShallowNet* \cite{schirrmeister2017}& cropped training &  $93.9$&-\\
     BaseNet & - &  $93.94 \pm 0.44$&$68.55\pm1.74$\\
     BaseNet + SE  & r=8  &   $94.45\pm0.5$& $67.53\pm1.62$\\
     \hline
     BaseNet + GSoP    &r=8 &   $93.66\pm0.87$ & $67.77\pm1.11$\\
     BaseNet + FCA    & r=8, $\text{DCT}^0$ & $94.17\pm0.51$ & $67.55\pm2.12$\\
     BaseNet + EncNet    &4 codewords &   $93.97\pm1.08$ & $67.29 \pm0.76$\\
     \hline
     BaseNet + ECANet  & k=15  & $\mathbf{94.88\pm0.85}$&  $67.4\pm1.95$\\
     \hline
     BaseNet + $\text{GE-}\theta^-$    & - &  $94.21\pm0.6$ & $68.21\pm1.29$\\
     BaseNet + $\text{GE-}\theta$    & e=global &   $94.3\pm0.42$ & $67.44\pm0.53$\\
     BaseNet + $\text{GE-}\theta^+$    & e=global, r=8 &   $93.77\pm1.162$ & $67.39\pm1.56$\\
     BaseNet + GCT    & - &   $94.19\pm0.53$ & $\mathbf{68.95\pm1.24}$\\
     BaseNet + $\text{GCT}_\text{GAP}$    & - &  $94.27\pm0.37$& $68.55\pm1.34$\\
     BaseNet + $\text{SE-}l_2$    &r=8 &   $94.62\pm0.35$ & $67.77\pm1.05$\\
     BaseNet + SRM  & -  & $93.3\pm0.39$ & $66.65\pm1.92$\\
     BaseNet + $\text{SRM}_{\text{cross}}$  & r=8  & $93.86\pm0.82$ & $66.13\pm1.11$\\
     \hline
     BaseNet + CBAM & r=4, k=7 & $94.77\pm0.4$ & $67.02\pm1.29$ \\
     BaseNet + CAT  & r=4, k=7 & $94.73\pm0.43$&  $67.24\pm1.98$\\
     BaseNet + CATLite & r=8   & $94.49\pm0.39$ &  $66.91\pm1.77$\\
     \hline
    TS-SEFFNet \cite{li2021temporal}& - &  $91.99\pm0.52$ & $69.99\pm0.57$\\
    EEGConformer \cite{song2022eeg}& - &  $\mathbf{95.32\pm0.4}$ & $70.63\pm1.28$\\
    ATCNet \cite{altaheri2022physics}& - &  $92.61\pm0.67$ & $65.95\pm1.51$ \\
    EEGTCNet \cite{ingolfsson2020tcn}& act=ELU &$92.24\pm1.2$ & $60.59\pm1.98$\\
     \hline
    \end{tabular}
\end{table}

\begin{table}[]
    \centering
    \caption{Results (in \%) of all models for the BCIC III IVa dataset in both settings along with their configuration.}
    \label{tab:results-bcic3}
    \begin{tabular}{ |p{4.5cm}||p{3cm}|p{3cm}|p{3cm}|  }
     \hline
     model & configuration & within-subject & cross-subject\\
     \hline \hline
     EEGNet \cite{lawhern2018eegnet}& - &   $71.39\pm1.59$& $64.96\pm1.37$\\
     ShallowNet \cite{schirrmeister2017}& - &  $72.64\pm0.77$&$68.32\pm1.57$\\
     BaseNet & - &  $76.82\pm1.25$&$65.82\pm2.93$\\
     BaseNet + SE  & r=1  &   $77.35\pm0.99$& $66.63\pm2.63$\\
     \hline
     BaseNet + GSoP    &r=1 &   $71.66\pm1.74$ & $66.21\pm1.8$\\
     BaseNet + FCA    & r=1, $\text{DCT}^0$ & $75.64\pm1.02$ & $\mathbf{66.81\pm2.51}$\\
     BaseNet + EncNet    &2 codewords &   $75.65 \pm 1.86$ & $66.71\pm2.16$\\
     \hline
     BaseNet + ECANet  & k=7  & $\mathbf{78.35\pm1.23}$&  $65.88\pm1.18$\\
     \hline
     BaseNet + $\text{GE-}\theta^-$    & - &  $77.31\pm0.97$ & $66.3\pm2.42$\\
     BaseNet + $\text{GE-}\theta$    & e=global &   $75.91\pm1.28$ & $66.72\pm1.99$\\
     BaseNet + $\text{GE-}\theta^+$    & e=global, r=1 &   $74.99\pm1.22$ & $66.69\pm0.84$\\
     BaseNet + GCT    & - &   $76.29\pm1.13$ & $65.76\pm2.87$\\
     BaseNet + $\text{GCT}_\text{GAP}$    & - &  $76.49\pm1.29$& $65.52\pm2.77$\\
     BaseNet + $\text{SE-}l_2$    &r=8 &   $77.29\pm0.98$ & $66.36\pm2.79$\\
     BaseNet + SRM  & -  & $77.26\pm0.92$ & $66.55\pm1.26$\\
     BaseNet + $\text{SRM}_{\text{cross}}$  & r=8  & $75.07\pm1.7$ & $66.04\pm1.74$\\
     \hline
     BaseNet + CBAM & r=4, k=3 & $77.08\pm0.91$ & $65.54\pm2.82$ \\
     BaseNet + CAT  & r=8, k=7 & $77.57\pm1.66$&  $66.61\pm1.33$\\
     BaseNet + CATLite & r=1   & $76.95\pm1.34$ &  $66.24\pm1.38$\\
     \hline
    EEGConformer \cite{song2022eeg}& - &  $\mathbf{81.71\pm0.54}$ & $\mathbf{73.37\pm0.58}$\\
    ATCNet \cite{altaheri2022physics}& - &  $72.16\pm1.02$ & $60.88\pm0.53$ \\
    EEGTCNet \cite{ingolfsson2020tcn}& act=ELU &$74.47\pm1.55$ & $62.23\pm1.65$\\
     \hline
    \end{tabular}
\end{table}

The development of BaseNet from ShallowNet and EEGNet is visualized in Figure \ref{fig:basent-development}. Developing BaseNet from ShallowNet, the introduction of the ELU function as well as the introduction of the depthwise separable convolution block along with the channel projector have the greatest impact on the performance. For the evolution from EEGNet to BaseNet on the other hand, changing the kernel and pool sizes shows the greatest impact. 
Developing BaseNet from ShallowNet and EEGNet, the performance improves with each design change except the simplification step.
We therefore also ran experiments without this simplification step. The final performance however was worse, indicating that those specialized design choices only suit their original architectures and training
routine but can easily be removed for BaseNet.\newline
In the upcoming sections, we will begin by presenting the results of the different channel attention mechanisms, along with the corresponding ablations. Next, we will assess these results against state-of-the-art approaches, both at the dataset level and on a per-subject basis. Finally, we will examine the models in terms of their computational cost.
\subsection{Channel attention mechanisms}
Table \ref{tab:results-bcic2a}-\ref{tab:results-bcic3} display the test accuracies of all tested models for all four datasets in both settings. The best model from the literature as well as our best model is indicated in bold. All results in the tables were produced by us except the ShallowNet* which is the reported result by \cite{schirrmeister2017} with their cropped training strategy. 
To investigate the design choices of the GCT block, we introduce two modifications $\text{GCT}_{\text{GAP}}$ and SE-$l_2$. $\text{GCT}_{\text{GAP}}$ replaces the original $l_2$ layer of GCT by a GAP layer, whereas SE-$l_2$ replaces the original GAP layer of SE by an $l_2$ layer. Additionally we introduce $\text{SRM}_\text{cross}$ where the original CFC layer is replaced by two FC layers and a ReLU function to allow cross-channel interaction.\newline
The hyperparameters reduction rate and kernel size were selected based on the ablations performed in section \ref{subsec:ablations}. The number of codewords of EncNet was set to the number of classes in the particular dataset. For FCA, our preliminary experiments showed that higher DCT components did not lead to any improvements and we therefore used the $\text{DCT}^0$ configuration across all experiments. This indicates that the element-wise multiplication with DCT components is not a powerful channel compression strategy for EEG decoding.
\subsubsection{BCIC IV 2a}
For the BCIC IV 2a dataset (Table \ref{tab:results-bcic2a}), CATLite, SE, CAT and ECANet showed the best results in the within-subject setting. For the cross-subject setting, CAT and EncNet performed well, but the overall improvements compared to BaseNet were smaller in contrast to the within-subject setting. 
Importantly, for the within-subject setting all attention modules performed at the level or above the level of BaseNet. Improving the squeeze module (GSoP, FCA, EncNet) resulted in a performance degradation compared to SE, while there was still an improvement compared to BaseNet.\newline
Among the solutions improving the squeeze and the excitation module, $\text{SRM}_\text{cross}$ showed the best results, indicating that cross-channel interaction is an important aspect of channel attention. As the results of GCT, $\text{GCT}_{\text{GAP}}$ and SE-$l_2$ are quite similar, we assume that both design choices of GCT have a similar impact on the performance. 
\newline 
CBAM, CAT and CATLite yield good results for both settings. This reflects that temporal attention as well as other compression strategies (maximum, entropy) can be useful. However, as CBAM and CAT do not outperform the other models systematically or by large margins, one could argue that temporal attenton is useful but not necessary.
\subsubsection{BCIC IV 2b}
For the BCIC IV 2b dataset (Table \ref{tab:results-bcic2b}) solutions without cross-channel interaction (GE-$\theta$, SRM) performed better than their cross-channel counterparts (GE-$\theta^+$, $\text{SRM}_\text{cross}$) in the within-subject setting. For the cross-subject setting, most approaches performed at the level or below the level of BaseNet. SE and CAT showed slightly better results than BaseNet. Apart from that, elaborating on additional observations is challenging as the differences between the attention modules are quite small.
\subsubsection{HGD}
For the HGD (Table \ref{tab:results-hgd}), ECANet is the best performing channel attention mechanism in the within-subject setting. The kernel size (k=15) is very large (compare Figure \ref{fig:eca-ablation}) which indicates the importance of a larger degree of cross-channel interaction.
For the cross-subject setting, the GCT module yields the best results. As $\text{GCT}_{\text{GAP}}$ performs better than SE and SE-$l_2$ the superior performance of GCT can be attributed to the gating mechanism rather than the $l_2$ layer.
\subsubsection{BCIC III IVa}
The ECANet shows the best results for the BCIC III IVa dataset (Table \ref{tab:results-bcic3}) in the within-subject scenario. The configuration (k=3), however, differs from the one used for the HGD dataset, indicating a lower degree of cross-channel interaction which is confirmed by the results of the different GE and SRM modules. For the cross-subject scenario the results of all attention modules are very similar and it is therefore difficult to make a statement about their performances.

\subsection{Ablations}
\label{subsec:ablations}
\begin{figure}
\centering
\begin{subfigure}{.41\textwidth}
    \centering
    \includegraphics[width=\textwidth]{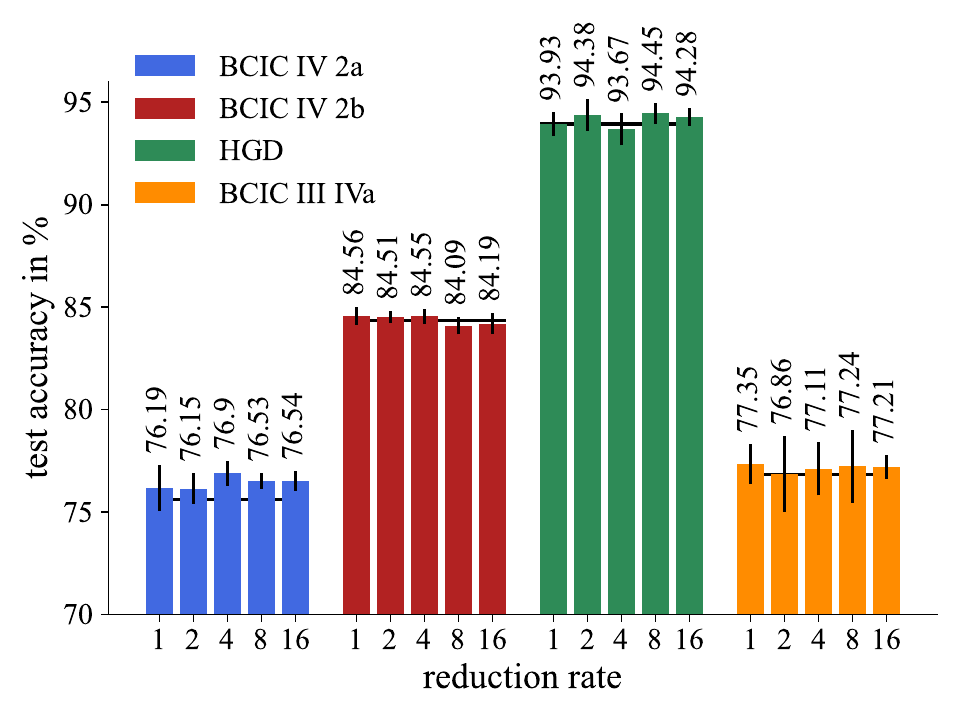}
    \caption{SENet}
    \label{fig:senet-ablation}
\end{subfigure}%
\hfill
\begin{subfigure}{.58\textwidth}
  \centering
    \includegraphics[width=\textwidth]{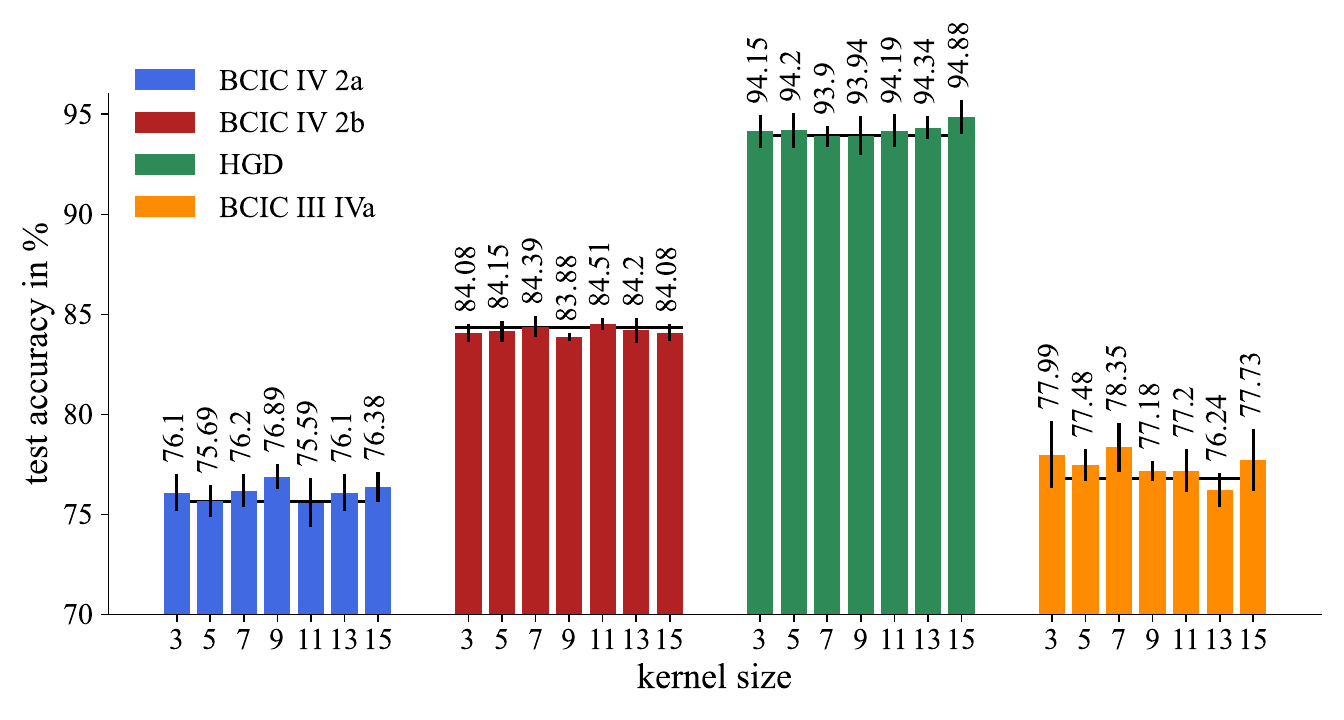}
    \caption{ECANet}
    \label{fig:eca-ablation}
\end{subfigure}
\caption{Ablation studies investigating the influence of the reduction rate and the kernel size for SENet and ECANet respectively. The black lines indicate the average test accuracy of BaseNet.}
\label{fig:sent-ecanet}
\end{figure}

\begin{figure}
\centering
\includegraphics[width=\textwidth]{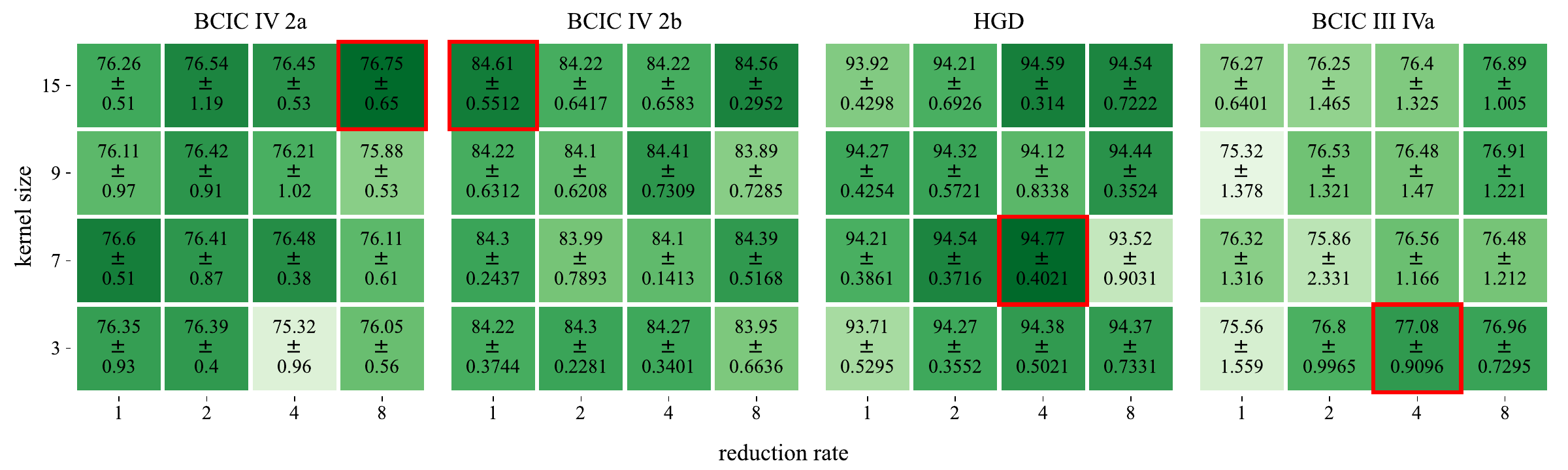}
\caption{Ablation studies investigating the relationship between the reduction rate $r$ in the channel attention module and the kernel size $k$ in the temporal attention module of CBAM. The best configuration is indicated by a red frame.}
\label{fig:cbam}
\end{figure}

\begin{figure}
\centering
\includegraphics[width=\textwidth]{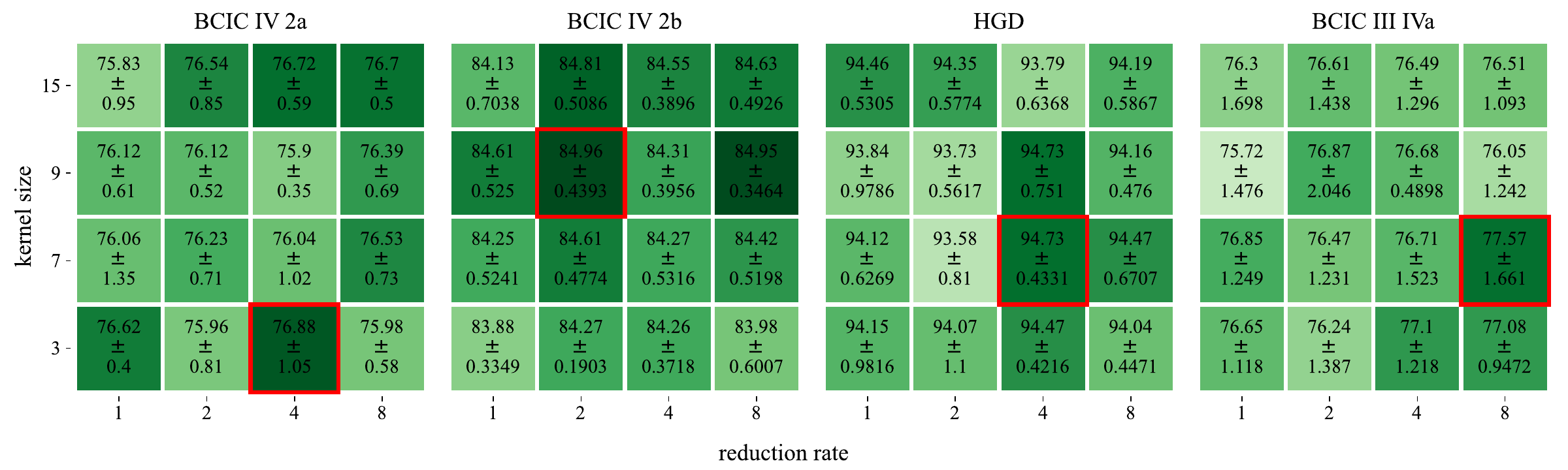}
\caption{Ablation studies investigating the relationship between the reduction rate $r$ in the channel attention module and the kernel size $k$ in the temporal attention module of CAT. The best configuration is indicated by a red frame.}
\label{fig:cat}
\end{figure}

To investigate the importance of the two most common hyperparameters reduction rate $r$ and kernel size $k$, we performed ablations for the within-subject setting for all four datasets for SE, ECANet, CBAM and CAT.\newline
The reduction rate $r$ determines the number of parameters ($2C^2/r$) and the degree of information reduction of an SE module for a given number of channels $C$. We evaluated different reduction rates in Figure \ref{fig:senet-ablation}.  For all datasets, the difference between the configurations is low but the SE module performs better than the BaseNet (indicated by the black line) for almost every configuration. For the BCIC IV 2a dataset, the improvement compared to the baseline is the largest. Additionally, every configuration works better than the BaseNet without attention.\newline 
To investigate the degree of cross-channel interaction, we varied the kernel size of ECANet and present the results in Figure \ref{fig:eca-ablation}. For the 2a dataset a medium size kernel works best, whereas a large kernel works best for the HGD. For the 2b dataset, the performance of ECANet is below or at the level of the baseline. In the BCIC III IVa dataset, the findings are varied, making it challenging to derive a clear observation.\newline
As CBAM and CAT both showed good performance but their best configurations regarding the reduction ratio $r$ and the kernel size $k$ differed, we decided to investigate this relationship closer. The results of these investigations are shown in Figure \ref{fig:cbam} and Figure \ref{fig:cat} for CBAM and CAT respectively.
The best configuration is marked with a red frame. Comparing all eight ablation studies, it is difficult to find signs that certain configurations generally work better than others. Interestingly, for the 2a dataset, the best configuration of CAT (r=4, k=3) marks the worst configuration for CBAM. For the HGD, on the other hand, the best configurations of CBAM and CAT match.


\subsection{Comparison with state of the art approaches}
We also compared our models with four state-of-the-art approaches for the same EEG motor imagery decoding task.
We included a lighweight architecture without attention (EEGTCNet), a model using the SE block (TS-SEFFNet) and two different approaches of using MHSA in the temporal domain to cover as many research directions as possible.
For EEGTCNet we used only a 4s window instead of the reported 4.5s window and the ELU activation function instead of ReLU, as it showed better results than their reported configuration. Apart from that, we used the same learning rate scheduler as in our experiments for TS-SEFFNet, EEGConformer and EEGTCNet as it also improved the results. All other hyperparameters were kept identical to the reported ones. It is worth noting that we train the EEGConformer twice as long (2000 epochs) and with their data augmentation method (segmentation and reconstruction) because the authors originally also trained the model like that.
As the BCIC III IVa dataset uses a smaller sampling rate than the other datasets, TS-SEFFNet, which employs wavelet convolutions specifically defined for a sampling rate of \SI{250}{\hertz}, could not be used for this dataset.\newline
Our simple BaseNet compares very well against more sophisticated and significantly larger (see number of paramters in Table \ref{tab:results-bcic2a}) architectures like TS-SEFFNet, EEGConformer, ATCNet and EEGTCNet. 
Although it rarely being the absolute best model, it consistently ranks among the highest-performing models across all datasets and settings. The strongest models out of the state-of-the-art models, namely EEGConformer and ATCNet on the other hand, perform very well in some settings at the cost of clearly underperforming in the remaining settings. EEGConformer for instance performs well for the BCIC IV 2a dataset in the within-subject setting and yields very good results for the HGD and BCIC III IVa dataset in both settings. For the remaining three settings on the other hand, it performs up to 13\% below BaseNet. For ATCNet, the opposite behavior is true: it performs very well for both BCIC IV datasets but clearly underperforms our models for the remaining two datasets.\newline
Compared to traditional DL approaches (ShallowNet, EEGNet), BaseNet outperforms EEGNet in all settings and ShallowNet in 6 out of 8 settings. Only for the cross-subject setting of the HGD and the BCIC III IVa dataset, ShallowNet outperforms BaseNet at the cost of underperforming it in other settings.
Across all datasets the channel attention modules are able to further improve the performance of BaseNet by a small margin.
\subsection{Results per subject}
\begin{figure}
\centering
\begin{subfigure}{.49\textwidth}
    \centering
    \includegraphics[width=\textwidth]{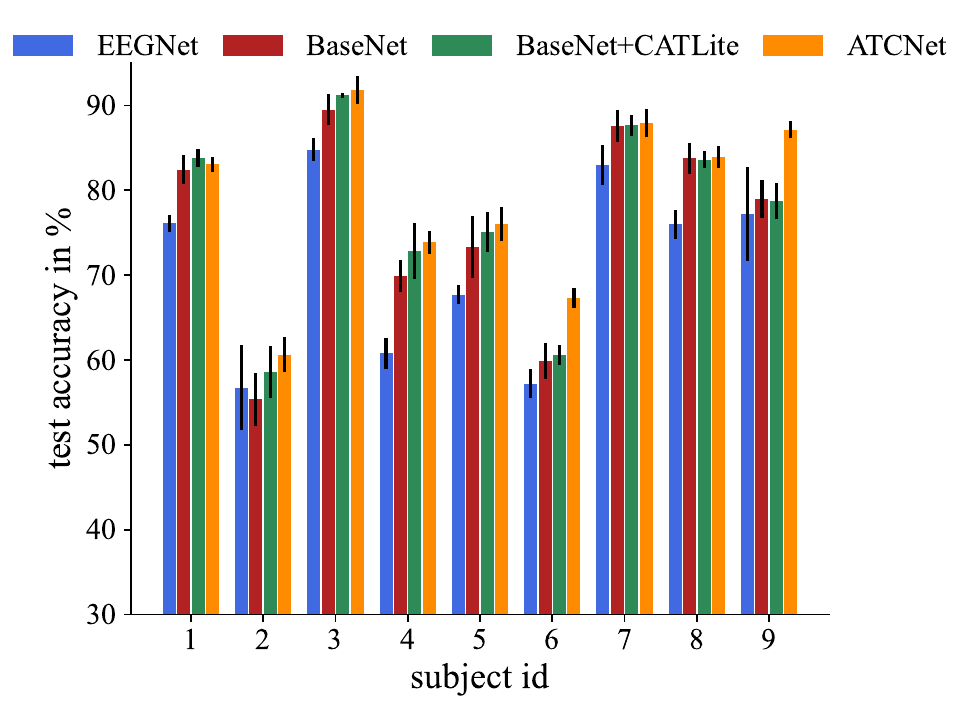}
    \caption{BCIC IV 2a}
    \label{fig:2a-per-subject}
\end{subfigure}%
\hfill
\begin{subfigure}{.49\textwidth}
  \centering
    \includegraphics[width=\textwidth]{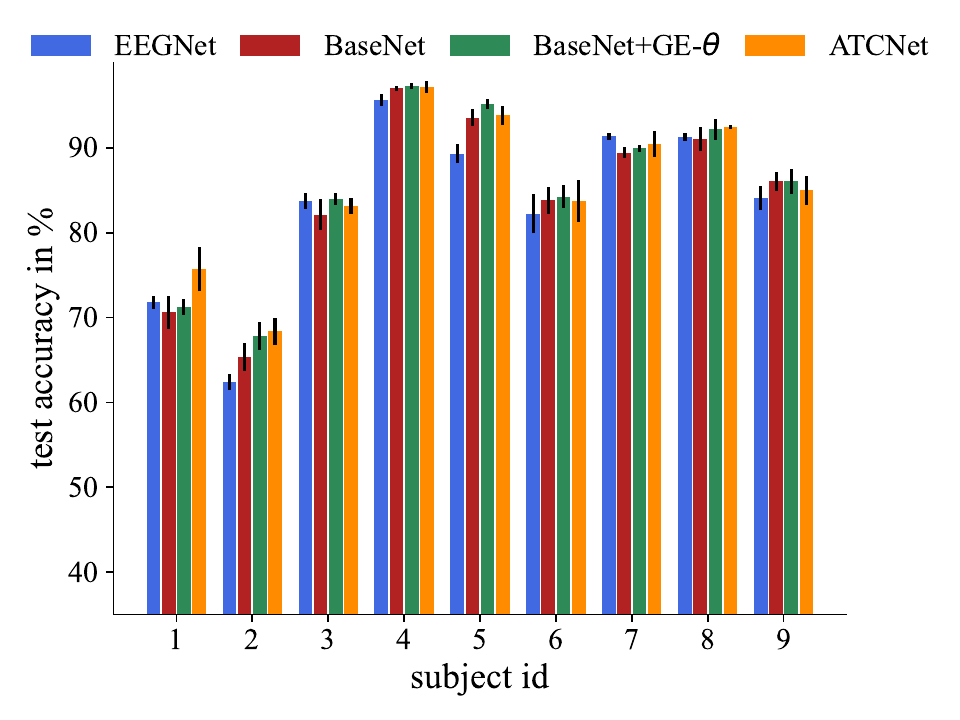}
    \caption{BCIC IV 2b}
    \label{fig:2b-per-subject}
\end{subfigure}
\vskip\baselineskip
\begin{subfigure}{0.7\textwidth}
        \centering
        \includegraphics[width=\textwidth]{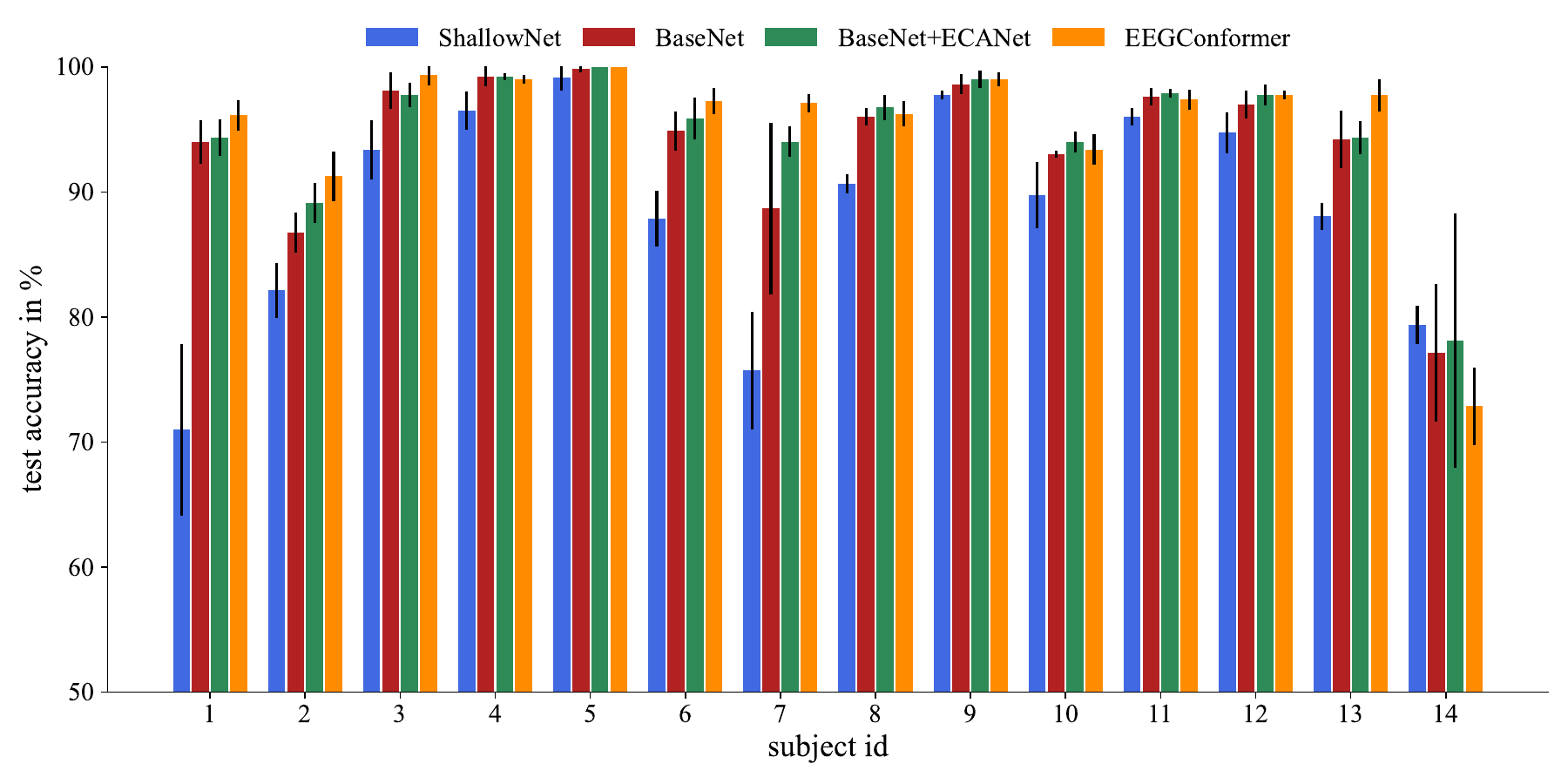}
        \caption{HGD}
        \label{fig:sub3}
\end{subfigure}
\hfill
\begin{subfigure}{0.29\textwidth}
        \centering
        \includegraphics[width=\textwidth]{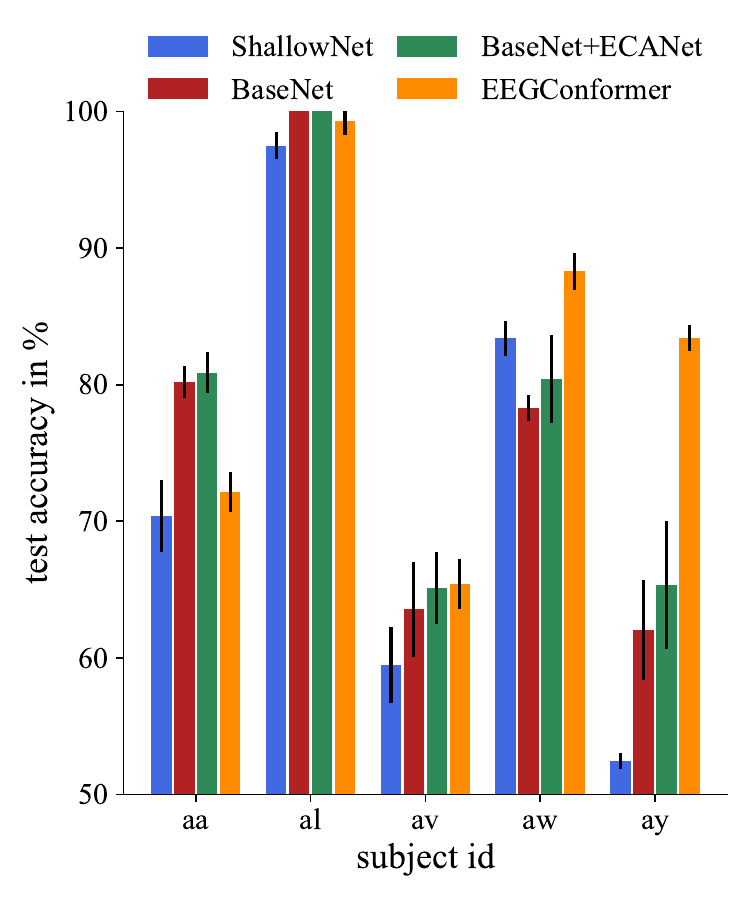}
        \caption{BCIC III IVa}
        \label{fig:sub3}
\end{subfigure}
\caption{Results per subject for all datasets in the within-subject setting. The black bar indicates the standard deviation between the five runs per model.}
\label{fig:results-per-subject-within}
\end{figure}

\begin{figure}
\centering
\begin{subfigure}{.49\textwidth}
    \centering
    \includegraphics[width=\textwidth]{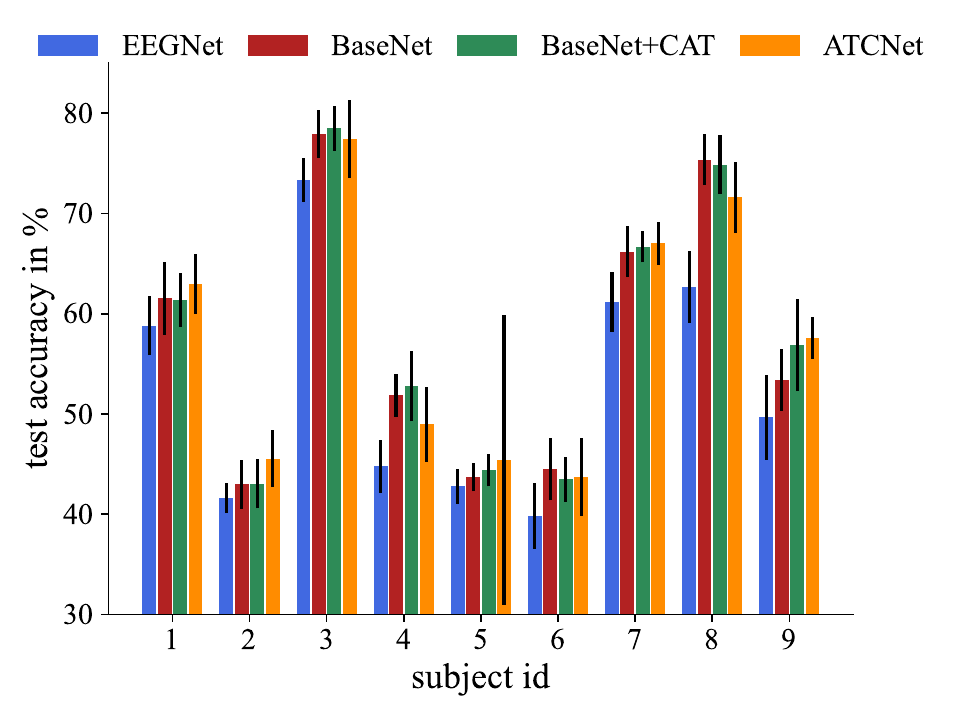}
    \caption{BCIC IV 2a}
    \label{fig:2a-per-subject}
\end{subfigure}%
\hfill
\begin{subfigure}{.49\textwidth}
  \centering
    \includegraphics[width=\textwidth]{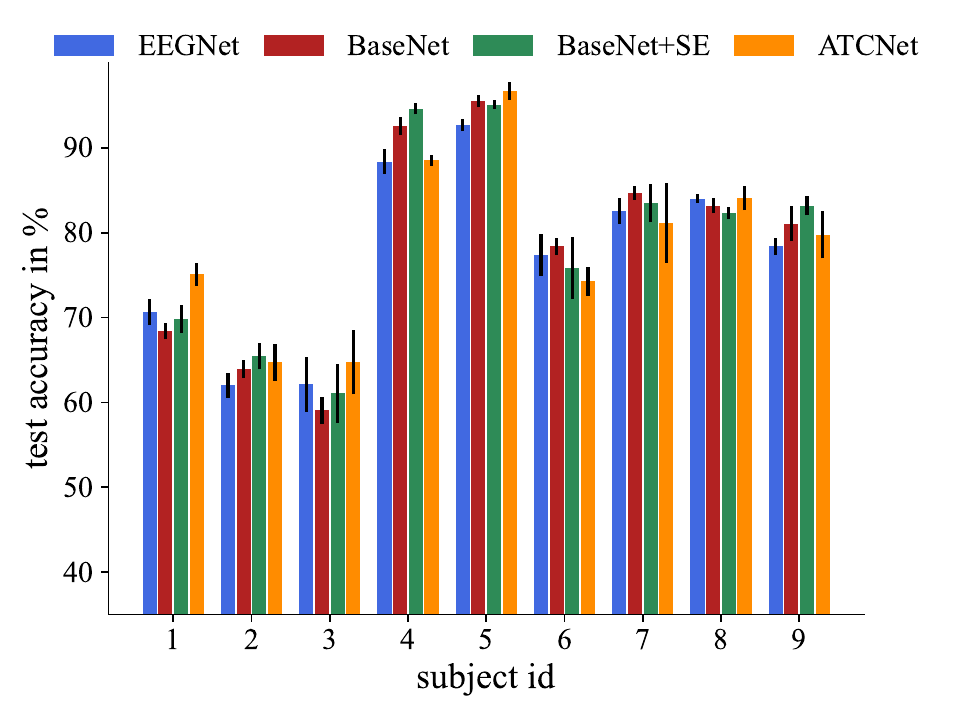}
    \caption{BCIC IV 2b}
    \label{fig:2b-per-subject}
\end{subfigure}
\vskip\baselineskip
\begin{subfigure}{0.7\textwidth}
        \centering
        \includegraphics[width=\textwidth]{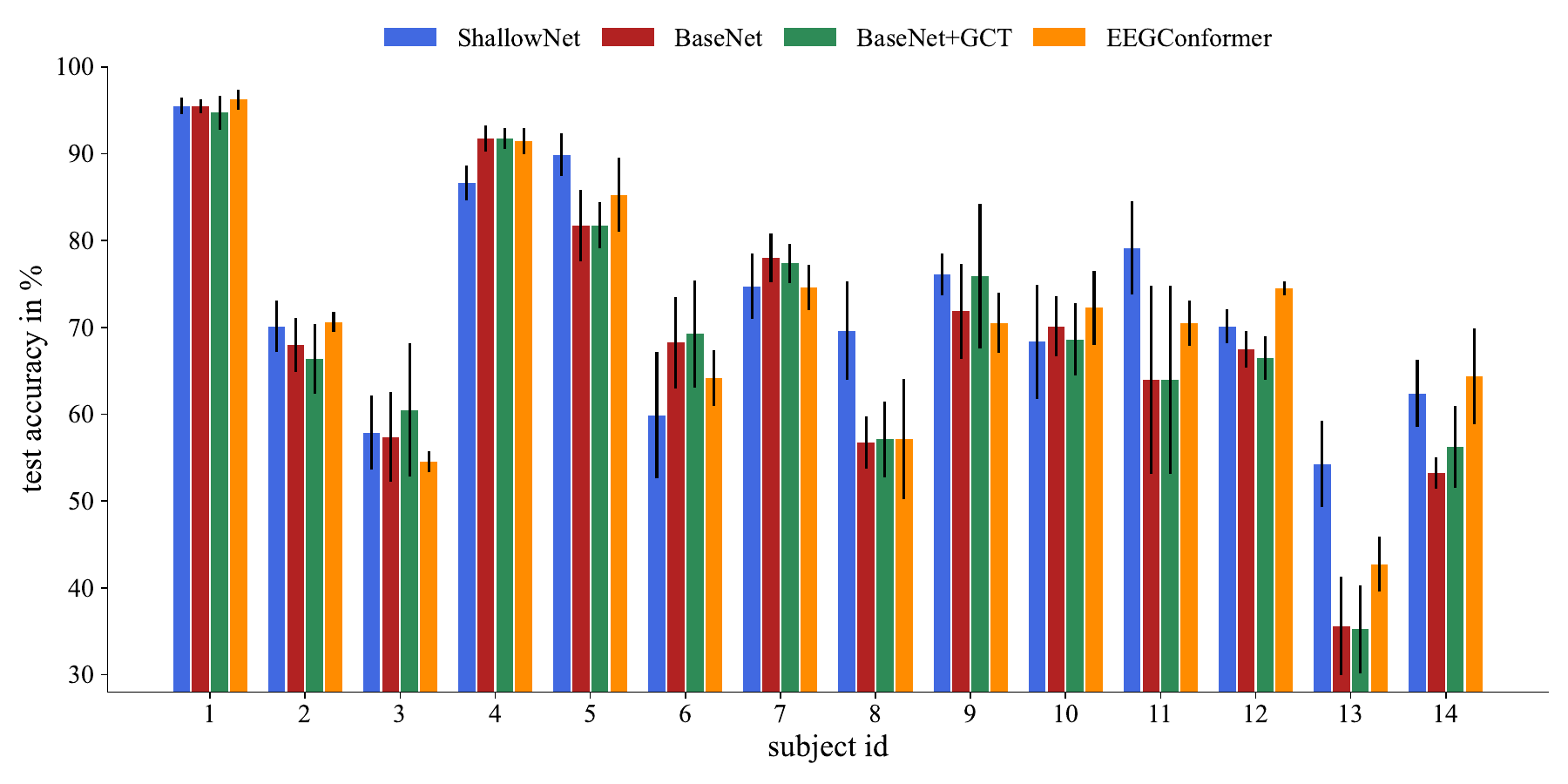}
        \caption{HGD}
        \label{fig:sub3}
\end{subfigure}
\hfill
\begin{subfigure}{0.29\textwidth}
        \centering
        \includegraphics[width=\textwidth]{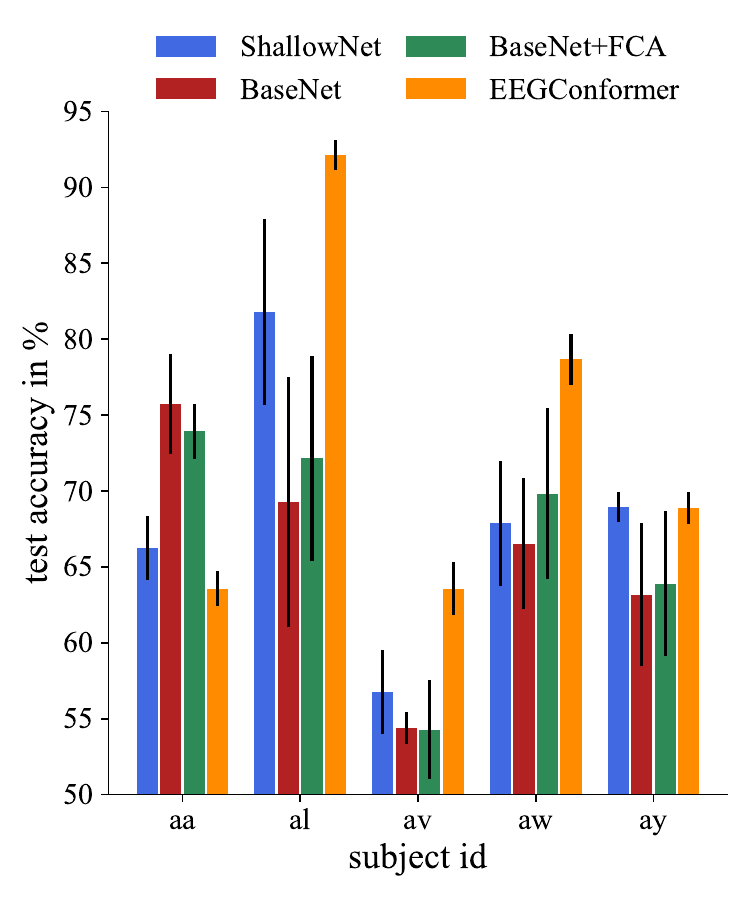}
        \caption{BCIC III IVa}
        \label{fig:sub3}
\end{subfigure}
\caption{Results per subject for all datasets in the cross-subject setting. The black bar indicates the standard deviation between the five runs per model.}
\label{fig:results-per-subject-loso}
\end{figure}

The Tables \ref{tab:results-bcic2a}-\ref{tab:results-bcic3} only display the mean performance averaged over all subjects. However, to assess the applicability and robustness of an algorithm, the performance per subject is also important. Therefore we additionally present the results of the best traditional DL model, BaseNet, the best model using channel attention and the best state-of-the-art model for the within-subject setting and the cross-subject setting in Figure \ref{fig:results-per-subject-within} and Figure \ref{fig:results-per-subject-loso}, respectively. \newline
For the BCIC IV 2a dataset in the within-subject setting, the overall differences are quite consistent with the improvements per subject. Subject 6 and subject 9 are the only clear outliers, where ATCNet outperforms the other models by a large margin. For the BCIC IV 2b dataset, there are no outliers and the improvements are distributed very evenly across all subjects.
For the HGD in the within-subject setting, only the results of ShallowNet exhibit irregularities as the performance is excessively low for subject 1 and subject 7 compared to the other models.
Due to the imbalanced number of training samples per subject in the BCIC III IVa dataset, the results between the subjects differ a lot from each other. The results for subject al (80-20 split) are almost perfect whereas the performance of subjects with less training data (av: 30-70 split, ay: 10-90 split) is significantly lower. Interestingly, the subject aw performs quite well compared to subject aa which has twice the amount of training data. Additionally, it is worth noting that the EEGConformer is able to achieve a significantly higher performance on the subject ay than all other models.\newline
For the cross-subject scenario, the BCIC IV 2a results are consistent. However, ATCNet exhibits a very large standard deviation between the runs for subject 5 which generally is among the most challenging subjects to decode. The results for the BCIC IV 2b dataset and the HGD display no irregularities. For the BCIC III IVa dataset, the results are very mixed and the differences between the models are only partially consistent with the overall differences. Surprisingly, the differences between the subjects are somewhat in line with the previous results from the within-subject scenario despite the significantly different distribution of training data.

\subsection{Computational cost} 
\begin{figure}
    \centering
    \includegraphics[width=0.8\textwidth]{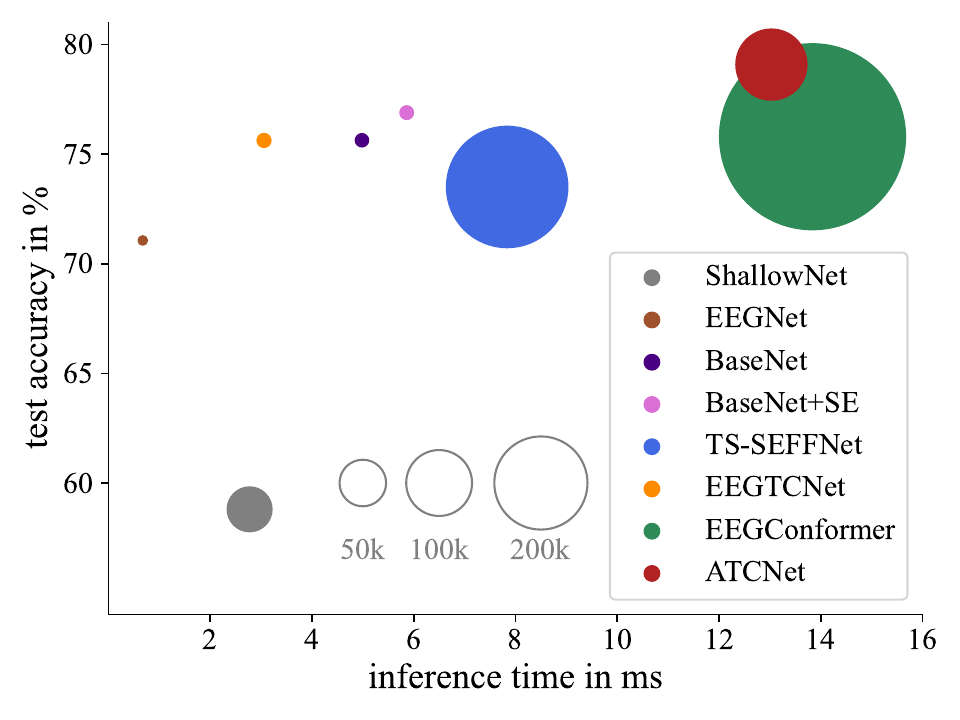}
    \caption{Inference time (measured on Intel i7-1195G7 with 4-cores), model size and test accuracy for the BCIC IV 2a dataset in the within-subject setting. The size of the circle indicates the number of parameters of each model. An ideal model would be given by a small dot in the upper left corner of the plot.}
    \label{fig:model-profiling}
\end{figure}

As BCI systems typically operate in online or closed-loop mode \cite{romero2020low, xu2013enhanced, mrachacz2016efficient, delisle2019system, stefano2022motor} on devices with limited computational resources, it is crucial to examine the computational expense of new algorithms. Clinicians tasked with developing online BCI systems with feedback must understand whether and how they can integrate advances in single-trial classification into their specific scenario.
Figure \ref{fig:model-profiling} displays the model size, inference time and the classification accuracy for the BCIC IV 2a dataset in the within-subject scenario. ATCNet and EEGConformer exhibit the longest inference times, while EEGNet demonstrates the lowest.  Our BaseNet typically operates with an inference time ranging from 5 to 6ms, contingent on the attention layer employed. In terms of model size or memory footprint, ShallowNet, ATCNet, TS-SEFFNet, and EEGConformer are the four largest models. Conversely, the remaining models, including the proposed BaseNet, have significantly reduced memory footprints, as evidenced by the smaller radii of their respective circles. \newline
Considering classification accuracy as well as inference time and memory footprint, three winners emerge: EEGTCNet, BaseNet, and BaseNet+SE. These models achieve high performance while maintaining a small memory footprint and reasonable inference time. However, for systems equipped with greater computational resources or those tolerant of larger inference times, ATCNet stands out as the optimal choice due to its superior performance.
Taking into consideration the classification accuracy shifts the ultimate selection based on the dataset and setting as discussed earlier. What remains constant is our framework's ability to deliver high performance while keeping memory usage small and inference time reasonable.
\newline
It's worth underlining that ATCNet utilizes a temporally stacked ensemble model, which is inefficient to optimize. Consequently, there is a 20x increase in training time compared to BaseNet with any attention layer (on a NVIDIA RTX A6000: 6 hours and 15 minutes versus 18.5 minutes for the 2a dataset). Depending on the available resources during training, this aspect should be taken into consideration.

\section{Discussion}
The results across all four datasets show that our proposed BaseNet is a very strong, yet lightweight, architecture that results in a good performance under different motor imagery decoding settings. The two strongest competitors EEGConformer and ATCNet produce the best results for some settings, at the cost of producing poor results on the remaining settings which limits their potential applicability to other datasets. As pointed out in the previous section, these two models have much larger computational demands during training and inference which further limits their applicability. So the lightweight architectures proposed by us might be the optimal choice if computational resources matter. In addition, our framework has better generalization capability across different datasets and settings. \newline
The results of the channel attention mechanisms along with the ablations show that while there is room for improvement, the improvements between the mechanisms and configurations are quite small compared to the differences due to major changes in the architecture (cf. Figure \ref{fig:basent-development}).
As we used the best configuration from the within-subject settings for the cross-subject settings, these configurations are suboptimal and we suppose that there is still some room for improvements. Larger improvements, however, can probably be achieved by tailoring the BaseNet architecture towards the datasets, e.g. by using more or less convolutional filters. The overall high performance across all datasets shows that our framework can be a solid starting point for new datasets and problems.
\newline
One surprising insight from our investigations is that the channel-independent SRM performed better than the cross-channel counterpart on the two datasets with only three sensors in the within-subject setting. On the 2a dataset with 22 sensors and the HGD with 44 sensors, however, the cross-channel version yields better results. This result indicates that the importance of cross-channel interaction might depend on the number of sensors.\newline
Using temporal and channel attention together, as in CBAM and CAT, resulted in good but not superior performance, suggesting that channel attention alone is sufficient and the combination with temporal attention is not necessary.
\subsection{Limitations}
To test the limits of our method, we also employed experiments under very challenging conditions with the BCIC III IVa dataset. The results were highly dependent on the subject and the amount of training data available. Compared to the other DL models applied in this study, our framework achieved good results in both settings. However, it is worth noting that the winners of the competition achieved far better results by employing traditional machine learning algorithms instead of deep learning. 
The results support our expectation that a carefully designed conventional machine learning algorithm still performs better for very small training datasets leveraging handcrafted feature engineering grounded in extensive domain expertise. Automated feature learning via a DL model on the other hand, requires more training data to become effective.\newline
Another limitation of our approach is the use of single-trial classification. Although our examination of memory footprint and inference time suggests that our framework may be suitable for real-time classification from a computational perspective, this aspect requires further investigation in future studies. It remains questionable how effectively the DL models can extract useful features from significantly shorter sequences which would be required for real-time applications.\newline
Additionally it is evident, that the classification accuracies for the cross-subject setting are substantially lower than for the within-subject setting across all datasets because of the distribution shift between training data and test data. This is expected and consistent across all models and datasets. These distributions shifts need to be addressed by domain adaptation methods to apply such models in practice.
\section{Conclusion}
We developed a simple and lightweight yet powerful framework for EEG motor imagery decoding which clearly outperforms the standard deep learning models EEGNet and ShallowNet. Further, it performs very well across all four datasets compared to more sophisticated, computationally demanding state of the art approaches. Additionally, we systematically investigated and compared a wide range of channel attention mechanisms which can be integrated seamlessly into our BaseNet while maintaining a low complexity and a small memory footprint. The results show that additional channel attention can further improve the performance of the proposed BaseNet.
\section*{Acknowledgments}
This paper has been generated within the BMBF-funded Quantum Human Machine Interfaces (QHMI) project, a component of the  QSens - Quantum Sensors of the Future cluster. We gratefully acknowledge the financial support provided by the BMBF, which was instrumental in advancing our research efforts. We also extend our deepest appreciation to our dedicated project partners, whose collaborative efforts were indispensable to the accomplishment of this study.
\section*{Conflict of Interest Statement}
The authors declare that the research was conducted in the
absence of any commercial or financial relationships that
could be construed as a potential conflict of interest.
\section*{References}
\bibliography{refs}
\bibliographystyle{ieeetr}
\end{document}